\documentstyle[preprint,aps,epsf]{revtex}
\tighten
\begin{document}
\draft
\title{Gravitational quasinormal modes for Anti-de Sitter black holes}
\author{Ian G Moss and James P Norman}
\address{
Department of Physics, University of Newcastle Upon Tyne, NE1 7RU U.K.
}
\date{December 2001}
\maketitle
\begin{abstract}
Quasinormal mode spectra for gravitational perturbations of black holes in
four dimensional de Sitter and anti-de Sitter space are investigated. The de
Sitter case lends itself to approximation by a P\"oshl-Teller potential. The
anti-de Sitter case is relevant to the ADS-CFT correspondence in superstring
theory. The ADS-CFT correspondence suggests a prefered set of boundary
conditions. \end{abstract}     
\narrowtext

\section{Introduction}

The spectrum of gravitational waves emitted by a black hole which has been
disturbed is dominated, over intermediate timescales, by characteristic
frequencies and decay timescales. These can be accurately described in terms
of the discrete set of quasinormal modes. The frequencies of these modes can
be determined numerically these days to high accuracy up to large order
\cite{nollert}. 

A new application of the quasinormal mode spectrum has arisen recently from
superstring theory. There is a suggestion, known as the ADS-CFT
correspondence, that string theory in anti-de Sitter space (usually with
extra internal dimensions) is equivalent to conformal field theory in one less
dimension \cite{maldacina,witten1}. A black hole, which has a characteristic
temperature fixed by the Hawking effect, should correspond to
CFT at finite temperature and the quasinormal modes would describe
non-equilibrium effects \cite{witten2}.

The temperature of the  black hole is fixed by the surface gravity of the
event horizon $\kappa_1$. The anti-de Sitter space also has a characteristic
length scale $a$ which is conventionally refered to as the anti-de Sitter
radius. In the non-rotating case, there are two black hole solutions for a
given value of the surface gravity \cite{page}, a larger one with positive and
smaller one with negative specific heat. Therefore the metric is determined by
$a$,  $\kappa_1$ and the choice of the larger or the smaller black hole.

Previous work on the quasinormal mode spectrum in anti-de Sitter space has
concentrated on the wave equation for scalar fields. Chan et al.
\cite{chan1,chan2} studied the scalar field problem prior to the interest in
the ADS-CFT correspondence. Horowitz et al. \cite{hor,hor2} examined the scalar
spectrum in four and five dimensions. They found simple scaling behaviour of
the spectrum in both limits $r_1\gg a$ and $r_1\ll a$, the former explained by
an approximate symmetry of the metric in that limit.

Gravitational perturbations of the BTZ black hole in three dimensions and
the anti-de Sitter black hole in four dimensions have been studied by Cardoso
et al. \cite{cardoso,cardoso2}. They make use of the fact that the
gravitational perturbation equation is separable in four dimensions 
\cite{mellor,guven,chambers}. This fact was also used in a preliminary report
on our results \cite{moss1}. The main difference between our work and the work
of Cardoso et al. is in the choice of boundary conditions on the axial
and polar metric perturbations.

There is a symmetry which relates the axial and polar metric perturbations,
whose existence can be explained by the fact that the covariant form of
perturbation analysis based on the Weyl tensor is governed by a single equation
\cite{chandra}. We argue that the thermal fluctuations in the energy of the
conformal field theory are determined by a single set of modes for which the
axial perturbations have dirichlet boundary conditions and the duality between
axial and polar perturbations is preserved. We calculate the quasinormal mode
frequencies numerically. For fluctuations in other components of the stress
energy tensor there is a one parameter family of boundary conditions and the
modes depend weakly on this parameter. 

We have also investigated the quasinormal mode spectrum for black holes in de
Sitter space. An interesting feature of this problem is that the asymptotic
behaviour of the modes is under better control than the corresponding problem
in flat space \cite{nollert}. The first few modes where evaluated previously
in ref. \cite{mellor}. Using methods based on continued fractions now gives us
many thousands of modes. A P\"oshl-Teller approximation for the
potential works extremely well in this example \cite{moss1}.

\section{De Sitter Space} 

The spherically symmetric de Sitter spacetime which contains a black hole has
two horizons, one at the event horizon of the black hole and a
cosmological horizon further out. In standard coordinates, the metric for a
black hole of mass $M$ and for a cosmological constant $\Lambda$ can be written
\begin{equation} ds^2=-{\Delta\over r^2}dt^2+{r^2\over\Delta}dr^2+
r^2(d\theta^2+\sin^2\theta\,d\phi^2)
\end{equation}
where
\begin{equation}
\Delta=r^2-2Mr-\case1/3\Lambda r^4.
\end{equation}
The event horizon lies at the smallest root $r_1$ of $\Delta$, where the
surface gravity $\kappa_1=r^{-2}\Delta'/2$. We can invert these relations to
give
\begin{eqnarray}
M&=&\case1/3 r_1(\kappa_1 r_1+1) \label{mass}\\
\Lambda&=&r_1^{-2}(1-2\kappa_1 r_1),\label{lambda}
\end{eqnarray}
showing, incidentally, that $\kappa_1 r_1<1/2$ in de Sitter space.

The gravitational wave equations were separated originally for charged black
holes \cite{mellor}, then for uncharged \cite{guven} and finally for rotating 
black holes \cite{chambers}. The metric perturbations can be classified by
their axial or polar symmetry under spatial inversions. The transformation
theory relating the two types is described in the next section. 

The axial metric perturbations with time dependence $e^{i\omega t}$ and
angular mode $l$ are subject to a one dimensional scattering equation,
\begin{equation}
{d^2 \psi\over dr^{*2}}+\left\{\omega^2-V(r)\right\}\psi=0\label{se}
\end{equation}
where
\begin{equation}
V(r)={\Delta\over r^4}\left( l(l+1)-{6M\over r}\right)\label{pot}
\end{equation}
The tortoise coordinate $r^*$, given as usual by $dr^*=r^2\,dr/\Delta$, covers
the range $-\infty<r^*<\infty$ as $r$ ranges between the two horizons. The
potential decays to zero exponentially fast in the $r^*$ coordinate at both
horizons. This is an important property which is not present for the black
hole in an asymptotically flat spacetime and guarantees that the reflection
and transmission coefficients are meromorphic functions of the frequency
\cite{newton}. 

By definition, the quasinormal modes are ingoing waves at the
event horizon and outgoing waves at the cosmological horizon,
\begin{equation}
\psi\to\cases{e^{-i\omega r^*}&as $r^*\to\infty$\cr
e^{i\omega r^*}&as $r^*\to-\infty$\cr}
\end{equation}
The method which we use to determine the complex frequencies follows the
continued fraction approach devised by Leaver \cite{leaver}. The idea is to
first scale out the divergent behavior at the cosmological horizon and examine
the resulting Fr\"obenius series. If we begin with a purely divergent solution
at the event horizon, the values of $\omega$ can be chosen to select the
convergent rescaled solution at the cosmological horizon.

The analysis is slightly easier if we use the independent variable $x=1/r$. In
de Sitter space, the radial equation has three regular singular points $x_1$
(the event horizon), $x_2$ (the cosmological horizon) and $x_3$. The
indices of the equation at these points are $\pm\rho_i=\pm
i\omega/(2\kappa_i)$, where 
\begin{eqnarray} 
\kappa_1&=&{2M(x_1-x_2)(x_1-x_3)}\\
\kappa_2&=&{2M(x_2-x_3)(x_2-x_1)}\\
\kappa_3&=&{2M(x_3-x_1)(x_3-x_2)}
\end{eqnarray}
Note that $\rho_1+\rho_2+\rho_3=0$ and
\begin{equation}
e^{-i\omega r^*}=(x-x_1)^{-\rho_1}(x-x_2)^{-\rho_2}(x-x_3)^{-\rho_3}
\end{equation}
We define,
\begin{equation}
\psi=e^{-i\omega r^*}u
\end{equation}
For a quasinormal mode, $u(x)$ must be divergent at $x_1$ and finite
at $x_2$. The Frobenius series is therefore
\begin{equation}
u(x)= (x-x_1)^{2\rho_1}\sum_{n=1}a_n\left({x-x_1\over x_2-x_1}\right)^n
\end{equation}
where $a_0=1$ and the coefficients satisfy a recurrence relation of the form
\begin{equation}
\alpha_n a_{n+1}+\beta_na_n+\gamma_na_{n-1}=0.\label{rec}
\end{equation}
The existence of this recurrence relation with only three terms is very
important, because the convergence of the Fr\"obenius series at $x=x_2$ is
now dependent on a continued fraction equation,
\begin{equation}
0=\beta_0-{\alpha_0\gamma_1\over \beta_1-}{\alpha_1\gamma_2\over \beta_2-}\dots
\end{equation}
which can be solved very quickly numerically.

\begin{figure}
\begin{center}
\leavevmode
\epsffile{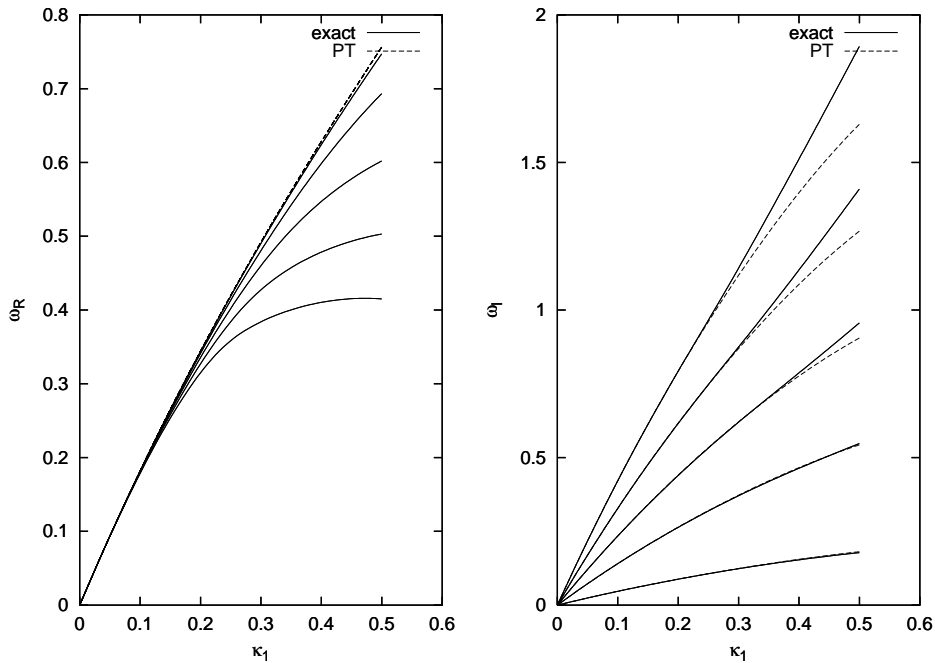}
\end{center}
\caption{Real (left) and imaginary (right) parts of the gravitational
quasinormal mode frequencies $r_1\omega$ with $l=2$ plotted as a function of
the surface gravity $r_1\kappa_1$, together with the P\"oschl-Teller
approximation.} \label{fig1}    
\end{figure}

\begin{figure}
\begin{center}
\leavevmode
\epsffile{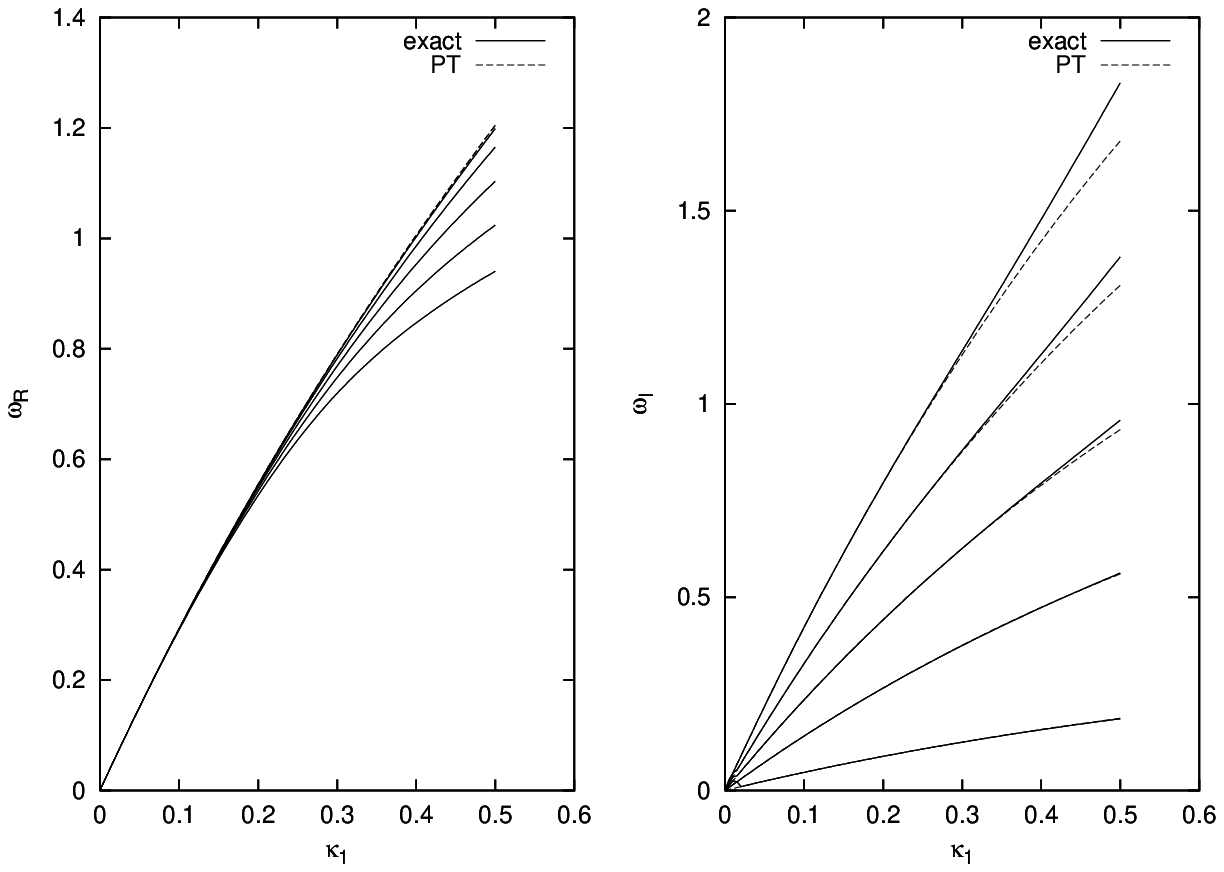}
\end{center}
\caption{Real (left) and imaginary (right) parts of the gravitational
quasinormal mode frequencies $r_1\omega$ with $l=3$ plotted as a function of
the surface gravity $r_1\kappa_1$, together with the P\"oschl-Teller
approximation.}  \label{fig2}   
\end{figure}

Figures \ref{fig1} and \ref{fig2} show the real and imaginary parts of
the frequency of the $l=2$ and $l=3$ modes as $r_1\kappa_1$ is varied. The
quasinormal modes in flat space are recovered at $r_1\kappa_1=0.5$.

The simple shape of the potential suggests that a P\"oschl-Teller
approximation might be effective, as it is for the Schwarzchild black hole
\cite{ferrari}. The approximate potential
\begin{equation}
V_{PT}=V_0\,{\rm sech}^2(r^*/b)
\end{equation}
contains two free parameters which are used to fit the height and the second
derivative of the potential at the maximum. The location of the maximum has
to be found numerically.

The quasinormal modes of the P\"oschl-Teller potential can be evaluated
analytically,
\begin{equation}
\omega_n={1\over b}\left\{(n+1/2)i+\sqrt{b^2V_0-1/4}\right\}.
\end{equation}
for $n=1,2,\dots$. These values are compared to the numerical values for $l=2$
in figure \ref{fig1} and $l=3$ in figure \ref{fig2}. The approximation works
best for the imaginary parts of the frequencies, but the differences between
the numerical values and the P\"oschl-Teller values decrease as $l$ is
increased and soon become smaller than the resolution of such graphs.

For large $l$ it is possible to obtain the P\"oschl-Teller fit to the
potential analytically. The maximum of the potential in this case lies close
to $r=3M$. From the second derivative at $r=3M$, we get
\begin{equation}
b^{-1}=\kappa_1(1+\kappa_1r_1)^{-1}(1+\case2/3\kappa_1r_1)^{1/2}
\end{equation}
and $b^2V_0=l(l+1)$. Equations (\ref{mass}) and (\ref{lambda}) have been
used to express the result in terms of the surface gravity. The analytic
expression for the frequencies is therefore
\begin{equation}
\omega_n=\kappa_1{(1+\case2/3\kappa_1r_1)^{1/2}\over 1+\kappa_1r_1}
\left\{(n+1/2)i+(l(l+1)-1/4)^{1/2}\right\}.\label{approx}
\end{equation}
Figures \ref{fig1} and \ref{fig2} show the numerical results and
the approximation for small $l$. The difference is largest in the flat space
limit at the right hand side of the figures. As $l$ is increased, the
difference beween the approximation and the numerical results soon becomes
insignificant.

\section{Anti-de Sitter space}

The metric for a black hole in anti-de Sitter space takes the same form as for
de Sitter space but with $\Lambda$ negative. The cosmological horizon is no
longer present. We introduce an anti-de Sitter radius $a$ by the definition
$a^2=-3/\Lambda$. For fixed values
of $\kappa_1$ and $a$ there are two black hole solutions with
different radii,
\begin{equation}
\kappa_1r_1=\case1/3(\kappa_1a)^2\pm\case1/3
\left((\kappa_1a)^4-3(\kappa_1a)^2\right)^{1/2}
\end{equation}  
The larger solution has positive and the smaller one negative specific heat.

The axial metric perturbations are subject to the same radial equation as
for de Sitter space, however in anti-de Sitter space the tortoise coordinate
lies in the range $-\infty<r^*\le 0$. The boundary conditions at $r^*=0$ can
be dirichlet, neumann or robin depending respectively on whether the field,
its derivative or a combination of both vanishes. Previous work has tended
to opt for dirichlet boundary conditions \cite{cardoso2}, based largely on
ADS invariance scalar field theory \cite{avis,breit}. However, dirichlet
boundary conditions break a duality in the equations between axial and
polar metric perturbations \cite{chandra}. We will introduce a set of self-dual
boundary conditions based on the ADS-CFT correspondence. 

In the Newman-Penrose formalism, the gravitational perturbations can be
described by two Weyl scalars $\Psi_0$ and $\Psi_4$. In flat space, $\Psi_0$
would represent gravitational waves which are incoming at infinity and
$\Psi_4$ would represent outgoing gravitational waves. The equations for the
Weyl scalars separate to give a generalisation of the Teukolski equations
\cite{chambers}. We put, for example, 
\begin{equation} 
\Psi_0=e^{i\omega
t}{r^3\over \Delta} Y(r)S_l(\theta) 
\end{equation}
where $S_l(\theta)$ can be found in Chandrasekhar \cite{chandra}. The radial
equation for $Y(r)$ can be related to the radial equation for the metric
perturbations (\ref{se}) if we set
\begin{equation}
Y=V^\pm\psi^\pm+(W^\pm+2i\omega)\Lambda_+\psi^\pm\label{np}
\end{equation}
where
\begin{equation}
\Lambda_+={d\over d r^*}+i\omega.
\end{equation}
There are two equivalent formulations corresponding to polar and axial
metric perturbations, denoted by the plus and minus signs. The axial form,
corresponding to $\psi^-$, has $V^-(r)=V(r)$ given previously (\ref{pot}).

The axial and polar functions are related by
\begin{equation}
C^\mp \psi^\mp=(C^\mp+72 M^2f)\psi^\pm
\mp 12M\Lambda_- \psi^\pm\label{trans}
\end{equation}
with $C^\pm=\mu^2(\mu^2+2)\pm12iM\omega$, and 
\begin{equation}
f={\Delta\over r^3}{1\over \mu^2r+6M}.
\end{equation}
where $\mu^2=(l-1)(l+2)$.

The ADS-CFT correspondence conjectures that when the Anti-de Sitter space is
conformally compactified, the generating functions of the conformal field
theory on the boudary are related to the partition function of the superstring
theory on the interior, or
\begin{equation} 
\left\langle 
\exp\left(\int \hat h_{\mu\nu}T^{\mu\nu}\right)
\right\rangle_{CFT}=
Z\left(\hat h_{\mu\nu}\right)\label{cor}
\end{equation}
If we calculate the correlation fuctions of a polar quantity, such as the
energy, we must therefore set the axial metric perturbation to vanish on the
boundary. The transformation theory (\ref{trans}) then implies a Robin boundary
condition on the polar metric perturbation, i.e. the normal derivative is
determined by the value of the field. The same set of quasinormal modes is
obtained from the axial and polar equations and the duality between the two is
preserved.

In general, both the axial and polar metric variations are non-vanishing.
This is consistent with the conformal theory viewpoint, where the six
components of the metric perturbations in three dimensions are subject to
three coordinate transformations and the conformal symmetry, leaving two
independent functions. If we define the ratio $\gamma=\psi^+/\psi^-$ , the two
metric variations are related by (\ref{trans}), which gives a robin boundary
condition 
\begin{equation}  
(\xi+\Lambda_-)\psi^-\label{gbc} 
\end{equation} 
where
\begin{equation}
\xi=(1-\gamma){C^+\over 12M}
+{6M\over\mu^2 a^2}
\end{equation}

Some special solutions lead to vanishing of the Newmann-Penrose scalars and
might be regarded as pure gauge modes from the point of view of the ADS theory.
For example, 
\begin{equation} \psi^-=e^{-i\omega r^*}\left(1+{6M\over
\mu^2r}\right) 
\end{equation} 
with $\omega=-\mu^2(\mu^2+2)i/12M$, obtained by setting $\psi^+=0$, is a purely
axial perturbation with $Y=0$. Replacing $\omega$ by $-\omega$ gives a
quasinormal mode with frequency 
\begin{equation}
\omega={\mu^2(\mu^2+2)\over 12M}i\label{pg}
\end{equation}
Such modes appear in the numerical solutions discussed below.

The frequencies of the quasinormal modes can be found numerically by
adapting the method used for de Sitter space. We first scale out the behaviour
of the quasinormal mode at the event horizon,
\begin{equation}
\psi^-=e^{i\omega r^*}u,
\end{equation}
where $u$ will be regarded as a function of $x=1/r$. The radial
equation (\ref{se}) gives the equation
\begin{equation} 
p\,u''+p'\,u'+2i\omega\,u'+\left\{l(l+1)-6Mx\right\}u=0
\end{equation} 
where $p=-a^{-2}-x^2+2Mx^3$. We take the Fr\"obenius series which is regular at
the event horizon,
\begin{equation}
u=\sum_{n=0}^\infty a_n\left({x-x_1\over -x_1}\right)^n
\end{equation}
with $a_0=1$. The coefficients are obtained from a recurrence relation similar
to (\ref{rec}). The series can then be substituted into the
boundary condition $u=0$ at $x=0$ to obtain an equation for the frequencies of
the quasinormal modes with vanishing axial perturbations.  

\begin{figure}
\begin{center}
\leavevmode
\epsfxsize=30pc
\epsffile{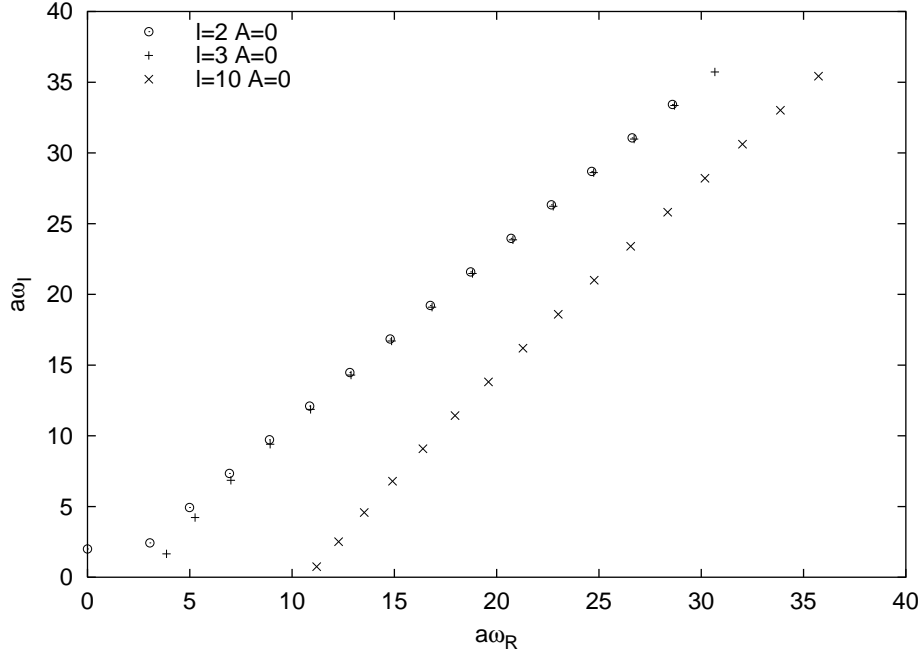}
\end{center}
\caption{The frequencies of the first few quasinormal modes for the
black hole anti-de Sitter metric with dirichlet axial boundary conditions
and $a\kappa_1=2.0$ (and $r_1=a$). Results for $l=2$, $3$ and $10$ are shown.}
\label{fig6}    \end{figure}

The frequencies of the first few modes using dirichlet boundary conditions
on the axial perturbations have been plotted for $a\kappa_1=2.0$ and $r_1=a$ in
figure \ref{fig6}. For $l=2$, one of the frequencies lies on the imaginary
axis. (This is also found by Cardoso et al \cite{cardoso}). We have found a
single frequency on the imaginary axis and can trace it for $\kappa_1r_1\ge
1.3$ when $l=2$, for $\kappa_1r_1\ge 3.4$ when $l=3$ and for $\kappa_1r_1\ge
38.0$ when $l=10$.

\begin{figure}
\begin{center}
\leavevmode
\epsffile{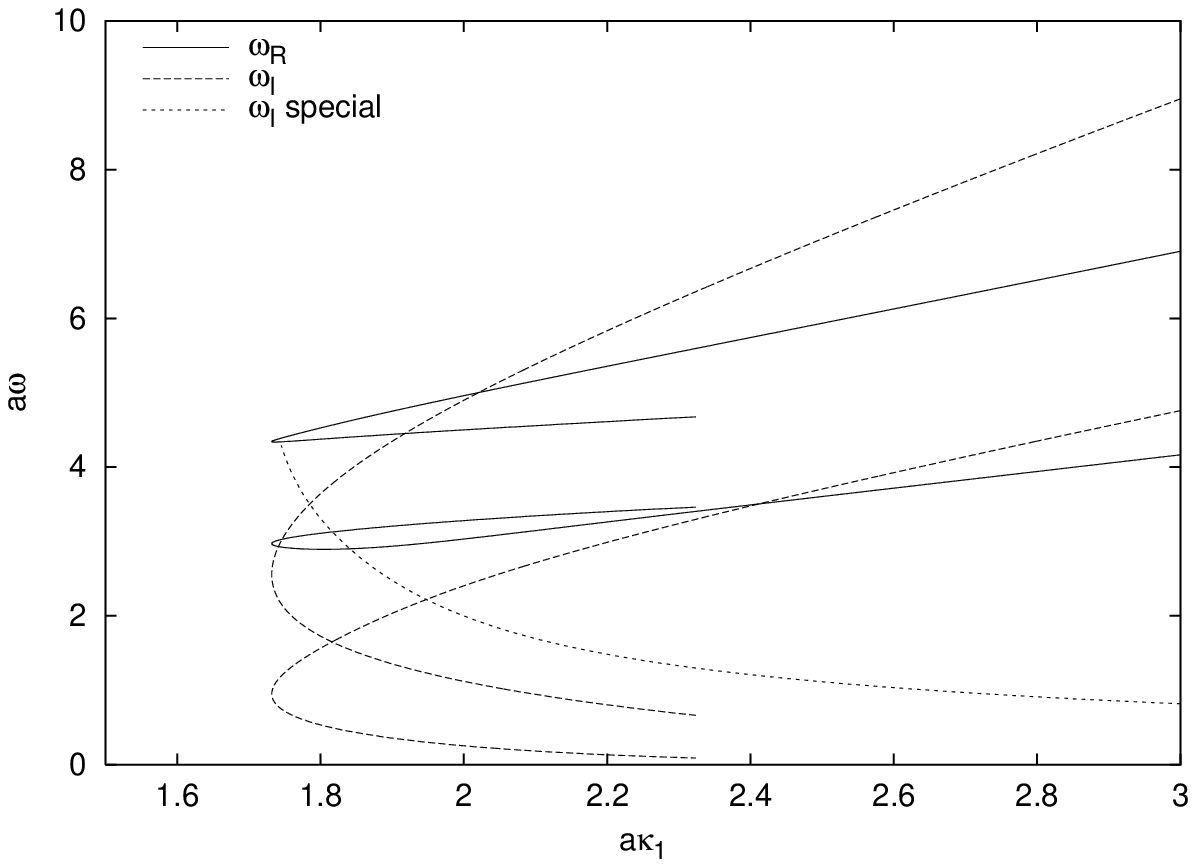}
\end{center}
\caption{The dependence of the first two quasinormal mode frequencies
on the metric parameters for $l=2$ is shown for the dirichlet axial boundary
conditions. The frequency with $\omega_R=0$ is also plotted and designated
`special'. } \label{fig7}     \end{figure}

The dependence of the first three modes on the metric parameters, with $l=2$
and $l=10$, is displayed in figures \ref{fig6} and \ref{fig7}. Both the real
and imaginary parts of the frequencies become proportional to $\kappa_1$ for
large $\kappa_1$, which agrees with the behaviour for the scalar field
quasinormal modes \cite{hor2}. For $r_1\ll a$, the numerical data suggests
that imaginary part of the frequency $\omega_I$ approaches zero as in the
scalar case \cite{hor2}. 

\begin{figure}
\begin{center}
\leavevmode
\epsfxsize=30pc
\epsffile{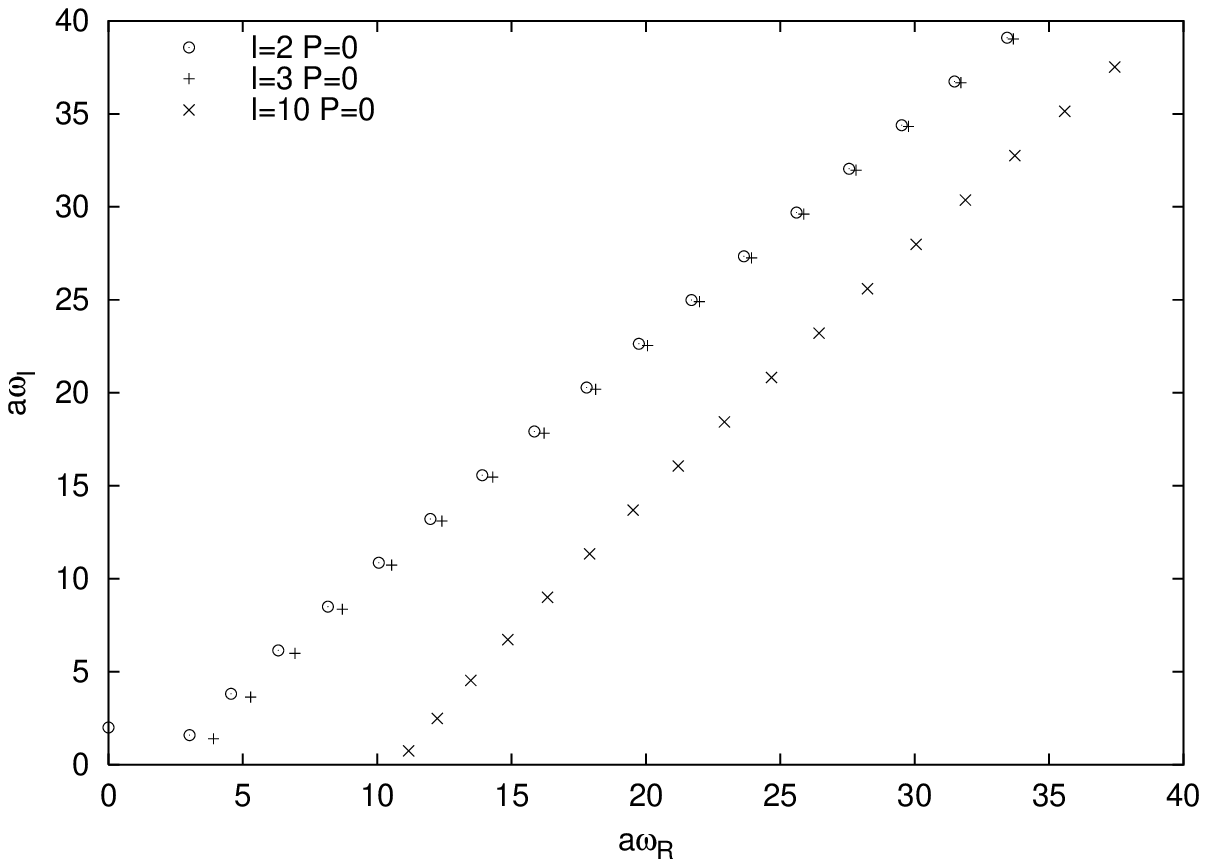}
\end{center}
\caption{The frequencies of the first few quasinormal modes for the
black hole anti-de Sitter metric with the dirichlet polar boundary conditions
($\gamma=0$) and $\kappa_1a=2.0$ (and $r_1=a$). Results for $l=2$, $3$ and
$10$ are shown.} \label{fig8}    \end{figure}

\begin{figure}
\begin{center}
\leavevmode
\epsffile{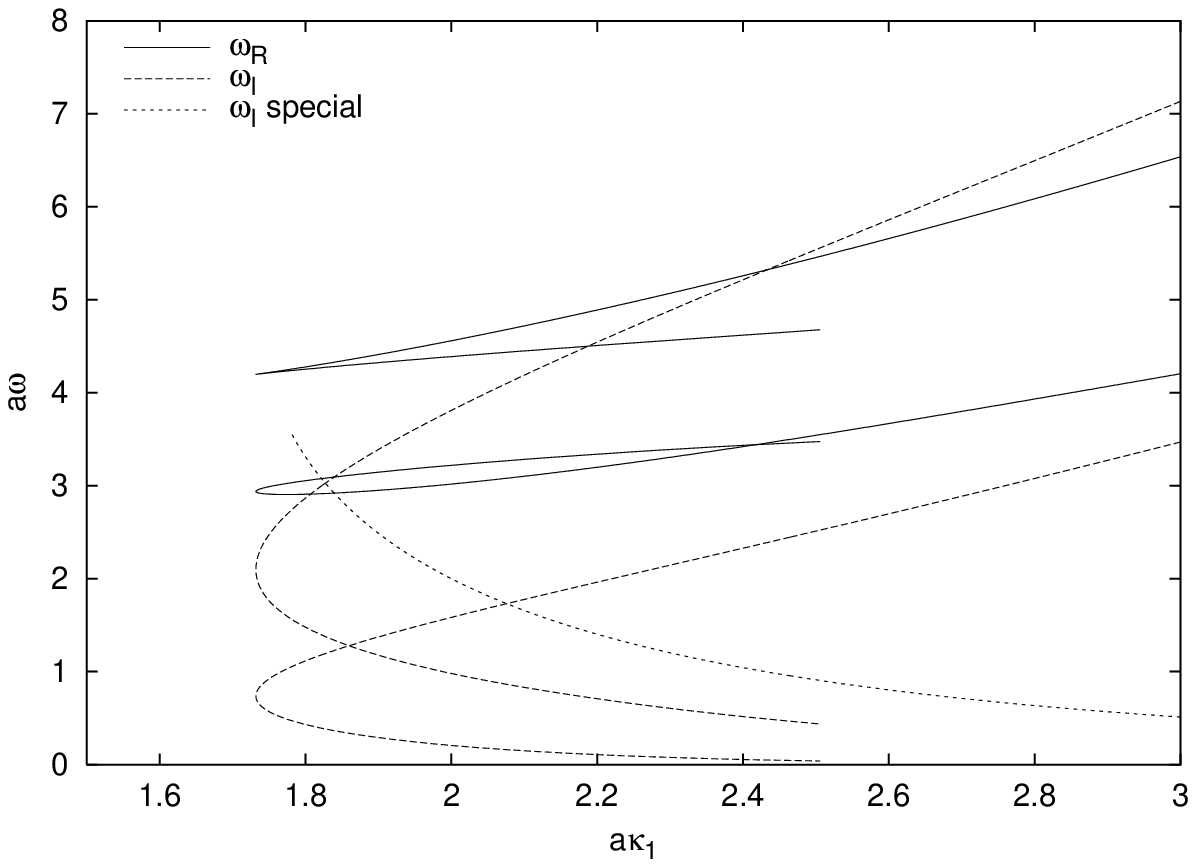}
\end{center}
\caption{The dependence of the first two quasinormal mode frequencies
on the metric parameters for $l=2$ is shown for the dirichlet polar boundary
conditions ($\gamma=0$). The frequency with $\omega_R=0$ is also plotted and
designated `special'. } \label{fig9}     \end{figure}

Corresponding results for dirichlet boundary conditions on the polar
perturbations ($\gamma=0$) are shown in figures \ref{fig8} and \ref{fig9}. The
special mode in this case is the `pure gauge' mode (\ref{pg}) and the analytic
solution provides a good check on the accuracy of our numerical technique. The
frequencies for these boundary conditions are not drastically different from
the dirichlet axial boundary conditions. The two sets of results represent two
extreme cases of the one parameter family of boundary conditions (\ref{gbc})
given above, and indicate a weak dependence on the parameter $\gamma$.

The potential $V(r)$ for anti-de Sitter space no longer vanishes as
$r\to\infty$ and it is harder to fit the potential to a P\"oschl-Teller form.
More importantly, the value of $r^*$ at infinity is zero, and the asymptotic
forms of the mode functions for the P\"oschl-Teller potential can no longer
be used.

\section{conclusion}

The quasinormal modes for the gravitational perturbations of a black hole in
de Sitter space have much in common with the quasinormal modes for the black
hole in asymptotically flat spacetime. The radial scattering problem in de
Sitter space is better behaved because the potential falls off exponentially
rather than as a power law. This explains why the approximation to the
quasinormal mode frequencies, derived in section 2 from a P\"oschl-Teller
potential, is in such good agreement with the numerical results for even
moderate angular order $l$.

The corresponding problem in anti-de Sitter space requires careful
consideration of the boundary conditions at infinity. The proposal here
has been to use the ADS-CFT correspondence and the duality symmetry between
axial and polar metric perturbations. For correlation functions of the energy
in the conformal field theory, the appropriate boundary conditions are that the
axial metric perturbations vanish and the polar perturbations satisfy a robin
boundary condition.

In general, two functions have to be specified on the boundary and there is a
one parameter family of boundary conditions. The effect on the quasinormal
frequencies is quite small.

It would be desirable to analyse thermal widths in three dimensional
conformal field theory to look for signs of a correspondence with the
quasinormal mode frequencies found here. This could give an impetus for
analysing the five dimensional gravitational quasinormal modes, where a full
separation of the gravitational perturbation equations is still an unsolved
problem.

\end{document}